\documentclass[final,leqno]{siamltex}

\def\draft{0}

\usepackage
{
        amssymb,
        amsmath,
        graphicx
}

\usepackage[margin=1in]{geometry}
\usepackage{amsfonts,latexsym,xspace,makeidx,bm,times}
\usepackage[dvipsnames,usenames]{color}
\usepackage[pagebackref,letterpaper=true,colorlinks=true,pdfpagemode=none,urlcolor=blue,linkcolor=blue,citecolor=BrickRed,pdfstartview=FitH]{hyperref}
\usepackage{boxedminipage}

\newcommand{\mynote}[2]{\marginpar{\tiny {\bf {#2}:} \sf {#1}}}

\ifnum\draft=1
\newcommand{\vnote}[1]{\mynote{#1}{VG}}
\newcommand{\aknote}[1]{\mynote{#1}{AKS}}
\newcommand{\ynote}[1]{\mynote{#1}{YZ}}
\else
\newcommand{\vnote}[1]{}
\newcommand{\aknote}[1]{}
\newcommand{\ynote}[1]{}
\fi

\newcommand{\eps}{\varepsilon}

\newcommand {\calC}   {{\cal{C}}}

\newcommand {\calE}   {{\cal{E}}}
\newcommand {\calH}   {{\cal{H}}}
\newcommand {\calL}   {{\cal{L}}}

\newtheorem{observation}[theorem]{Observation}
\definecolor{DSgray}{cmyk}{0,0,0,0.7}

\newcommand{\R}{\mathbb R}

\usepackage{prettyref}

\newcommand{\savehyperref}[2]{\texorpdfstring{\hyperref[#1]{#2}}{#2}}

\newrefformat{eq}{\savehyperref{#1}{\textup{(\ref*{#1})}}}
\newrefformat{lem}{\savehyperref{#1}{Lemma~\ref*{#1}}}
\newrefformat{def}{\savehyperref{#1}{Definition~\ref*{#1}}}
\newrefformat{thm}{\savehyperref{#1}{Theorem~\ref*{#1}}}
\newrefformat{cor}{\savehyperref{#1}{Corollary~\ref*{#1}}}
\newrefformat{corr}{\savehyperref{#1}{Corollary~\ref*{#1}}}
\newrefformat{cha}{\savehyperref{#1}{Chapter~\ref*{#1}}}
\newrefformat{sec}{\savehyperref{#1}{Section~\ref*{#1}}}
\newrefformat{app}{\savehyperref{#1}{Appendix~\ref*{#1}}}
\newrefformat{tab}{\savehyperref{#1}{Table~\ref*{#1}}}
\newrefformat{fig}{\savehyperref{#1}{Figure~\ref*{#1}}}
\newrefformat{hyp}{\savehyperref{#1}{Hypothesis~\ref*{#1}}}
\newrefformat{alg}{\savehyperref{#1}{Algorithm~\ref*{#1}}}
\newrefformat{item}{\savehyperref{#1}{Item~\ref*{#1}}}
\newrefformat{step}{\savehyperref{#1}{step~\ref*{#1}}}
\newrefformat{conj}{\savehyperref{#1}{Conjecture~\ref*{#1}}}
\newrefformat{fact}{\savehyperref{#1}{Fact~\ref*{#1}}}
\newrefformat{prop}{\savehyperref{#1}{Proposition~\ref*{#1}}}
\newrefformat{claim}{\savehyperref{#1}{Claim~\ref*{#1}}}
\newrefformat{obs}{\savehyperref{#1}{Observation~\ref*{#1}}}
\let\pref=\prettyref

\newcommand{\Sref}[1]{\hyperref[#1]{Section \ref*{#1}}}

\newcommand{\bU}{{\boldsymbol{U}}}
\newcommand{\bW}{{\boldsymbol{W}}}
\newcommand{\bbU}{{\overline{\boldsymbol{U}}}}

\newcommand{\E}{\mathop{\bf E\/}}

\renewcommand{\phi}{\varphi}

\newcommand{\norm}[1]{\left\lVert #1 \right\rVert}

\newcommand{\balsep}{{\sc BalancedSeparator}\xspace}
\newcommand{\scut}{{\sc UniformSparsestCut}\xspace}
\newcommand{\threexor}{{\sc 3-XOR}\xspace}
\newcommand{\maxcut}{{\sc MaxCut}\xspace}

\newcommand{\edges}{{\rm edges}}
\newcommand{\bad}{{\rm bad}}

\title{{Constant Factor Lasserre Integrality Gaps for Graph Partitioning Problems}}

\author{Venkatesan Guruswami\thanks{Computer Science Department, Carnegie Mellon University. Supported in part by a Packard Fellowship and NSF CCF-1115525. Email: \texttt{guruswami@cmu.edu}} \and Ali Kemal Sinop\thanks{Computer Science Department, Princeton University. Supported by NSF CCF-1115525 and MSR-CMU Center for Computational Thinking. Email: \texttt{asinop@cs.cmu.edu}} \and Yuan Zhou\thanks{Computer Science Department, Carnegie Mellon University. Supported in part by NSF CCF-1115525 and US-Israel BSF grant 2008293. Email: \texttt{yuanzhou@cs.cmu.edu}}}

\begin{document}
\maketitle
\begin{abstract}
Partitioning the vertices of a graph into two roughly equal parts while minimizing the number of edges crossing the cut is a fundamental problem (called Balanced Separator) that arises in many settings. For this problem, and variants such as the Uniform Sparsest Cut problem where the goal is to minimize the fraction of pairs on opposite sides of the cut that are connected by an edge, there are large gaps between the known approximation algorithms and non-approximability results. While no constant factor approximation algorithms are known, even APX-hardness is not known either.

In this work we prove that for balanced separator and uniform sparsest cut, semidefinite programs from the Lasserre hierarchy (which are the most powerful relaxations studied in the literature) have an integrality gap bounded away from $1$, even for $\Omega(n)$ levels of the hierarchy. This complements recent algorithmic results in Guruswami and Sinop (2011) which used the Lasserre hierarchy to give an approximation scheme for these problems (with runtime depending on the spectrum of the graph). Along the way, we make an observation that simplifies the task of lifting ``polynomial constraints" (such as the global balance constraint in balanced separator) to higher levels of the Lasserre hierarchy.
\end{abstract}

%\thispagestyle{empty}
%\end{titlepage}

%\tableofcontents
%\newpage



\pagestyle{myheadings}
\thispagestyle{plain}
\markboth{GURUSWAMI ET AL.}{LASSERRE GAPS FOR GRAPH PARTITIONING PROBLEMS}

\allowdisplaybreaks

\section{Introduction}

Partitioning a graph into two (balanced) parts with few edges going across them is a fundamental optimization problem. Graph partitions or separators are widely used in many applications (such as clustering, divide and conquer algorithms, VLSI layout, etc). Two prototypical objectives of graph partitioning are \balsep~ and \scut, defined as follows.
\begin{definition}
Given an undirected graph $G= (V, E)$ and $0 < \tau < 0.5$, the goal of the $\tau$ vs $1-\tau$ \balsep~ problem is to find a set $A \subseteq V$ such that $\tau |V| \leq |A| \leq (1 - \tau) |V|$, while $\edges(A, V \setminus A)$ is minimized. Here $\edges(A, B)$ is the number of edges in $E$ that cross the cut $(A, B)$.

The goal of the \scut~ problem is to find a set $\emptyset \subsetneq A \subsetneq V$ such that the sparsity
\begin{align*}
\frac{\edges(A, V \setminus A)}{|A| |V \setminus A|}
\end{align*}
is minimized.
\end{definition}

Despite extensive research, there are still huge gaps between the known approximation algorithms and inapproximability results for these problems. The best algorithms, based on semidefinite relaxations (SDPs) with triangle inequalities, give a $O(\sqrt{\log n})$ approximation to both problems \cite{ARV09}. On the inapproximability side, a Polynomial Time Approximation Scheme (PTAS) is ruled out for both problems assuming {\sc 3-SAT}~does not have randomized subexponential-time algorithms \cite{AMS11}. 
In this paper, our focus is on the \scut problem; the general {\sc SparsestCut} problem has been shown to not admit a constant-factor approximation algorithm under the Unique Games Conjecture~\cite{CKKRS06, KV05, Kho02}.

It is known that the SDP used in \cite{ARV09} cannot give a constant factor approximation for \scut~\cite{DKSV06}. Integrality gaps are also known for stronger SDPs: super-constant factor integrality gaps for both \balsep~and \scut~are known for the so-called Sherali-Adams$_+$ hierarchy for a super-constant number of rounds~\cite{RS09b}.
There has been much recent interest in the power and limitations of various {\em hierarchies} of relaxations in the quest for better approximation algorithms for combinatorial optimization problems. These hierarchies are parameterized by an integer $r$ (called rounds/levels) which capture higher order correlations between (roughly $r$-tuples of) variables (the basic SDP captures only pairwise correlations, and certain extensions like triange inqualities pose constraints on triples).  Larger the $r$, tighter the relaxation.
% guaranteed to find an optimal integral solution. 

There are several hierarchies of relaxations that have been studied in the literature, such as Sherali-Adams~\cite{SA90},  Lov\'{a}sz-Schrijver hierarchy~\cite{LS91}, ``mixed" hierarchies combining Sherali-Adams linear programs with SDPs, and the {\em Lasserre hierarchy}~\cite{Las02}. Of these hierarchies, the most powerful one is the Lasserre hierarchy (see \cite{laurent-comparison} for a comparison). The potential of SDPs from the Lasserre hierarchy in delivering substantial improvements to approximation guarantees for several notorious optimization problems is not well understood, and is an important and active direction of current research. Indeed, it is consistent with current knowledge that even 4 rounds of the Lasserre hierarchy could improve the factor $0.878$ Goemans-Williamson algorithm for \maxcut, and therefore refute the Unique Games conjecture. Very recent work \cite{BBH+12, OZ13, DMN13, KOTZ14} has shown that $O(1)$ levels of the Lasserre hierarchy can succeed where $\omega(1)$ levels of weaker SDP hierarchies fail; in particular, this holds for the hardest known instances of {\sc UniqueGames} \cite{BBH+12}.

For the graph partitioning problems of interest in this paper (\balsep~and \scut), integrality gaps were not known even for a small constant number of rounds. It was not (unconditionally) ruled out, for example, that $1/\epsilon^{O(1)}$ rounds of the hierarchy could give a $(1+\epsilon)$-approximation algorithm, thereby giving a PTAS. On the algorithmic side, \cite{GS11} recently showed that for these problems, SDPs using $O(r/\eps^2)$ rounds of the Lasserre hierarchy have an integrality gap at most $(1 + \epsilon)/\min\{1,\lambda_r\}$.
Here $\lambda_r$ is the $r$-th smallest eigenvalue of the normalized Laplacian of the graph. This result implies  an approximation scheme for these problems with runtime parameterized by the graph spectrum.

Given this situation, it is natural to study the limitations of the Lasserre hierarchy for these two fundamental graph partitioning problems. Several of the known results on strong integrality gap results for many rounds of the Lasserre hierarchy, starting with Schoenebeck's remarkable construction~\cite{Sch08}, apply in situations where a corresponding NP-hardness result is already known. Thus they are not ``prescriptive" of hardness. In fact, we are aware of only the following examples where a polynomial-round Lasserre integrality gap stronger than the corresponding NP-hardness result is known: Max $k$-CSP, $k$-coloring \cite{Tul09} and Densest $k$-Subgraph \cite{BCGVZ11}. The main results of this paper, described next, extend this body of results, by showing that Lasserre SDPs cannot give a PTAS for \balsep~and \scut.

%
%\ynote{Venkat: if you are going to remove this sentence, please be aware that the part about maxcut results also needs to be revised correspondingly.}
%

%Several limitations of the Lasserre hierarchy are known, but for most of them we already know the corresponding NP-Hardness.

\subsection{Our results}
In this paper, we study integrality gaps for the Lasserre SDP relaxations for \balsep~and \scut. As mentioned before, APX-hardness is not known for these two problems, even assuming the Unique Games conjecture. (Superconstant hardness results are known based on a strong intractability assumption concerning the Small Set Expansion problem~\cite{RST10}.) In contrast, we show that linear-round Lasserre SDP has an integrality gap bounded away from $1$, and thus fails to give a factor $\alpha$-approximation for some absolute constant $\alpha > 1$. Specifically, we prove the following two theorems.

\begin{theorem}[informal] \label{thm:informal-balsep}
For $0.45 < \tau < 0.5$, there are linear-round Lasserre gap instances for the $\tau$ vs $(1-\tau)$ \balsep~ problem, such that the integral optimal solution is at least $(1+\epsilon(\tau))$ times the SDP solution, where $\epsilon(\tau) > 0$ is a constant dependent on $\tau$.
\end{theorem}

\begin{theorem}[informal] \label{thm:informal-scut}
There are linear-round Lasserre gap instances for the \scut~problem, such that the integral optimal solution is at least $(1+\epsilon)$ times the SDP solution, for some constant $\epsilon > 0$.
\end{theorem}

\subsection{Our techniques}

All of our gap results are based on Schoenebeck's ingenious Lasserre integrality gap for \threexor\ \cite{Sch08}. For \balsep~and \scut, we use the ideas in \cite{AMS11} to build gadget reductions and combine them with Schoenebeck's gap instance. \cite{AMS11} designed gadget reductions from Khot's quasi-random PCP \cite{Kho06} in order to show APX-Hardness of the two problems. If we view the Lasserre hierarchy as a computational model (as suggested in \cite{Tul09}), we can view Schoenebeck's construction as playing the role of a quasi-random PCP  in the Lasserre model.
%(the structure is slightly different from the one in \cite{Kho06}).
Our gadget reductions, therefore, bear some resemblance to the ones in \cite{AMS11}, though the analysis is different due to different random structures of the PCPs. We feel our reductions are slightly simpler than the ones in \cite{AMS11}, although we need some additional tricks to make the reductions have only linear blowup. This latter feature is needed in order to get Lasserre SDP gaps for a linear number of rounds. We are also able to make the gap instance graphs have only {\bf constant degree}, while the reductions in \cite{AMS11} give graphs with unbounded degree.

Also, unlike 3-XOR, for balanced separator there is a global linear constraint (stipulating the balance of the cut), and our Lasserre solution must also satisfy a lifted form of this constraint~\cite{Las02}. We make a general observation that such constraints can be easily lifted to the Lasserre hierarchy when the vectors in our construction satisfy a related linear constraint. This observation applies to constraints given by any polynomials, and to our knowledge, was not made before. It simplifies the task of constructing legal Lasserre vectors in such cases.

\section{Lasserre SDPs}

In this section, we begin with a general description of semidefinite programming relaxations from the Lasserre hierarchy, followed by a useful observation about constructing feasible solutions for such a SDP. We then discuss the specific SDP relaxations for our problems of interest. Finally, we recall Schoenebeck's Lasserre integrality gaps~\cite{Sch08} in a form convenient for our later use.

\subsection{Lasserre Hierarchy Relaxation}
%The Lasserre Hierarchies for Graph Partitioning Problems}
%
%Given an undirected graph $G = (V, E)$, we want to divide the vertex set into two parts : $A$ and $V \setminus A$. The $r$-round Lasserre SDP relaxation introduces a vector $\bbU_{S}$ for each subset $S \subseteq V$ with $|S| \leq r$. The inner product of the vectors $\langle \bbU_{S_1}, \bbU_{S_2} \rangle$ can be viewed as the probability that $S_1 \cup S_2 \subseteq A$, where in integral solutions, this value is either $0$ or $1$. The Lasserre constraints insist the following consistency checks of the inner products (which we refer to as \emph{general Lasserre conditions})
%%
%%\begin{itemize}
%%%\item $\langle \bbU_{S_1}, \bbU_{S_2}\rangle \geq 0$ for all $S_1, S_2$;
%%\item $\langle \bbU_{A}, \bbU_{B} \rangle = \|\bbU_{A\cup B}\|^2$;
%%\item  $\norm{\bbU_{\emptyset}}^2 = 1$.
%%\end{itemize}

\let\es=\emptyset

%%\begin{definition}[Moment Matrix]
%%Given positive integers $n$ and $r$, $r^{th}$ round moment matrix
%%is a linear function $\mathbb{M}_r: \R^{\binom{[n]}{\le 2 r}} \to \R^{
%%\binom{[n]}{\le r} \times \binom{[n]}{\le r}}$ over matrices
%%whose rows and columns are associated with subsets of size $\le r$, whose
%%entry at row $A_1 \in \binom{[n]}{\le r}$ and column $A_2 \in \binom{[n]}{\le r}$
%%is given by the following:
%%\[
%%\mathbb{M}_r(y)_{A_1, A_2} \triangleq y_{A_1 \cup A_2}.
%%\]
%%\end{definition}
%%\begin{notation}[Convolution]
%%Given two vectors $P \in \R^{\binom{[n]}{\le d}}$ and $y \in \R^{\binom{[n]}{\le r}}$ , their convolution $(P\ast y)\mapsto \R^{\binom{[n]}{\le r-d}}$ is the following vector valued bilinear function. For any $T \in \R^{\binom{[n]}{\le r}}$:
%%\[
%%(P\ast y)_{T} = \sum_{S} P_S y_{S\cup T}.
%%\]
%%\end{notation}
Consider a binary programming problem with polynomial objective function $P$ and a single constraint
expressed as a polynomial $Q$:
\begin{equation}\label{eq:ip-000}
\begin{array}{rll}
\text{Minimize/Maximize } & \sum_{T\in\binom{[n]}{\le d}} P(T) \prod_{j\in T} x_j \\
\text{subject to } &
%\sum_{j\ge 1} \left(\sum_{T\in\binom{[n]}{\le d}} Q_{j}(T) \prod_{j\in T}
%x_j\right)^2 \le Q_0,\\
\sum_{T\in\binom{[n]}{\le d}} Q(T) \prod_{j\in T} x_j \ge 0,\\
& x_i \in \{0,1\} \qquad \text{ for all $i\in [n]$.}
\end{array}
\end{equation}
It is easy to see that this captures all problems we consider in this paper:
\balsep~(\pref{sec:balsep}) and \scut~(\pref{sec:scut}).

We now define the Lasserre hierarchy semidefinite program relaxation
for the above integer program. It is easily seen that the below is a
relaxation by taking $\bbU_A = x_A \mathbf{I}$ and $Y_A = \sqrt{Q(x)}\
\bbU_A$ where $x \in \{0,1\}^n$ is a feasible solution to
\eqref{eq:ip-000}, $x_A = \prod_{i\in A} x_i$, and $\mathbf{I}$ is any
fixed unit vector.
\begin{proposition}
  For any positive integer $r \ge d$, $r$ rounds of Lasserre Hierarchy
  relaxation~\cite{Las02} of \pref{eq:ip-000} is given by the
  following semidefinite programming formulation:
  \begin{equation}\label{eq:sdp-000}
    \begin{array}{rll}
      \text{Minimize/Maximize } & \sum_{T} P(T) \left\| \bbU_{T}\right\|^2 \\
      \text{subject to } &
      \|\bbU_{\es}\|^2 = 1, \\
      & \langle \bbU_{A}, \bbU_B\rangle = \|\bbU_{A\cup B}\|^2 \qquad \text{ for all $A, B$ with $|A\cup B|\le 2 r$}, \\
      % & Q_0 \|\bbU_{A\cup B} \|^2 - \sum_{j\ge 1} \bigg\langle \sum_{S} Q_{j}
      % (S) \bbU_{S\cup A}, \sum_{T} Q_{j}(T) \bbU_{T\cup B} \bigg\rangle = \|
      % Y_{A\cup B}\|^2 \\
      & \sum_{S\in\binom{[n]}{\le d}} Q(S) \langle
      \bbU_{S} ,
      \bbU_{A\cup B}
      \rangle
      = \langle Y_A, Y_B\rangle,\\
      % & \qquad\qquad\qquad
      % \qquad\qquad\qquad \text{ for all $A, B$ with $|A\cup B|\le 2 (r-d)$,}\\
      & \langle Y_A, Y_B\rangle = \|Y_{A\cup B}\|^2
      \qquad\quad\text{ for all $A, B$ with $|A\cup B|\le 2 (r-d)$.}\\
      & \bbU_A,Y_B \in \R^{\Upsilon}.
    \end{array}
  \end{equation}
\end{proposition}
\def\mmt{\mathbb{M}}
\begin{proof}
  Given $y \in \R^{\binom{[n]}{\le 2 r}}$, let $\mmt(y)\in
  \mathsf{Sym}(\binom{[n]}{\le r})$ be the moment matrix whose rows
  and columns correspond to subsets of size $\le r$. The entry at row
  $S$ and column $T$ of $\mmt(y)$ is given by $y_{S \cup T}$. For any
  multilinear polynomial $P$ of degree-$d$, let $P \ast y \in
  \R^{\binom{[n]}{\le 2 r - d}}$ be the vector whose entry
  corresponding to subset $S$ is given by $\sum_{T} P_T y_{S\cup T}$.
  The Lasserre Hierarchy relaxation~\cite{Las02} of \pref{eq:ip-000} 
  is given by:
  \begin{equation}\label{eq:sdp-639}
    \begin{array}{rll}
      \text{Minimize/Maximize } & \sum_T P(T) y_T  \\
      \text{subject to }       &y_{\es} = 1, \\
      &\mmt(y) \succeq 0, \\      
      &      \mmt(Q \ast y) \succeq 0.
    \end{array}
  \end{equation}

  {\it Proof of \pref{eq:sdp-000}$\implies$ \pref{eq:sdp-639}.}
  %{\bf Proof of \pref{eq:sdp-000}$\implies$ \pref{eq:sdp-639}}
    Given feasible solution for \pref{eq:sdp-000}, let $y_{S}
    \triangleq \|\bbU_S\|^2$ and $z_S \triangleq \|Y_S\|^2$. We have
    $y_{\es} = 1$ and $\sum_T P(T) y_T = \sum_T P(T) \|\bbU_S\|^2$.
    Observe that $y_{S\cup T} = \|\bbU_{S\cup T}\|^2 = \langle \bbU_S
    , \bbU_T\rangle$ therefore $\mmt(y) \succeq 0$. With a similar
    reasoning, we also have $\mmt(z)\succeq 0$. Finally, for any $S$:
  \[
  (Q \ast y)_S = \sum_{T} Q(T) y_{S\cup T} = \sum_T 
  Q(T) \langle
  \bbU_{T} ,
  \bbU_{S}
  \rangle
  = \|Y_{S}\|^2 = z_S,
  \] which implies $z = Q \ast y$. Hence $y$ is a feasible solution
  for \pref{eq:sdp-639}. 
  
  {\it Proof of \pref{eq:sdp-639}$\implies$ \pref{eq:sdp-000}.} Let $y$
  be a feasible solution for \pref{eq:sdp-639}. Define $z \triangleq Q
  \ast y$. Since $\mmt(y)\succeq 0$ (resp. $\mmt(z)\succeq 0$), there
  exists a matrix $\bbU=[\bbU_S]_S$ (resp. $Y=[Y_S]_S$) such that
  $\mmt(y) = \bbU^T \bbU$ (resp. $\mmt(z) = Y^T Y$).  It is easy to
  see that $\langle \bbU_S , \bbU_T \rangle = y_{S\cup T}$ and
  $\langle Y_S, Y_T \rangle = z_{S\cup T}$. Therefore:
  \begin{itemize}
  \item $\sum_T P(T) \|\bbU_T\|^2 = \sum_T P(T) y_T$.
  \item $\|\bbU_{\es}\|^2 = y_{\es} = 1$.
  \item $\langle \bbU_S , \bbU_T \rangle = y_{S\cup T} = \|\bbU_{S\cup
      T}\|^2$ (similar for $Y$).
  \item For any $S$, $\sum_T Q(T) \langle \bbU_T , \bbU_S \rangle
    = \sum_T Q(T) y_{S\cup T} = (Q \ast y)_S = z_S = \|Y_S\|^2$.
  \end{itemize}
  Therefore $(\bbU, Y)$ is a feasible solution for \pref{eq:sdp-000}
  with same objective value, completing our proof.
\end{proof}

% \begin{proof}
%   \aknote{Is it necessary? What do you guys think?}  \vnote{If it is
%   not too painful, perhaps we can recap what Lasserre says in terms
%   of moment/Gram matrices, and hopefully the translation to vector
%   form above is clear. If we want to handwave we can have it in text
%   before the lemma statement and not call it a proof :)}
% \end{proof}
Note that a straightforward verification of last two constraints
requires the construction of vectors $Y_A$ in addition to
$\bbU_A$. Below we give an easier way to verify these last two
constraints without having to construct $Y_A$'s. This greatly
simplifies our task of constructing Lasserre vectors for the lifting
of global balance constraints. % (which is of sum-of-squares type).
\begin{theorem} \label{thm:las-check}
Given vectors $\bbU_T$ for all $T\in \binom{[n]}{\le 2 r}$ satisfying the first two constraints
of \pref{eq:sdp-000},
if there exists a non-negative real $\delta>0$ such that
%if the following conditions hold then
%these vectors form (part of) a feasible solution: There exists a non-negative
%real $\delta>0$ such that:
%there exists vectors $Y$ satisfying
%the final two constraints of \pref{eq:sdp-000}:
%\begin{align}
%\forall j: \bigg\|\sum_{S} Q_{j}(S) \bbU_{S}\bigg\|^2= &
%\bigg( \sum_{S} Q_{j}(S) \big\|\bbU_{S}\big\|^2 \bigg)^2,
%  \label{eq:sdp-001}\\
%\sum_{j\ge 1} \bigg\|\sum_{S} Q_{j}(S) \bbU_{S}\bigg\|^2 \le & Q_0.\label{eq:sdp-002}
%\end{align}
\begin{align}
\sum_{S\in\binom{[n]}{\le d}} Q(S) \bbU_{S} = \delta \cdot \bbU_{\es}
\end{align}
then
these vectors form (part of) a feasible solution to \pref{eq:sdp-000}.
\end{theorem}
\begin{proof}
Consider the following vectors. For each $A$ with $|A|\le r$, let
$Y_A = \sqrt{\delta} \cdot \bbU_{A}$.
By construction, these vectors satisfy the $ \langle Y_A, Y_B\rangle = \|Y_{A\cup B}\|^2$ constraints since  $\langle \bbU_{A}, \bbU_B\rangle = \|\bbU_{A\cup B}\|^2$. Now we verify the  other constraint:
\begin{align*}
\sum_{S \in {[n] \choose \leq d}} Q(S) \langle \bbU_{S}, \bbU_{A \cup B} \rangle = \Bigg \langle \sum_{S \in {[n] \choose \leq d}} Q(S) \bbU_{S} , \bbU_{A \cup B} \Bigg \rangle = \langle \delta \bbU_{\emptyset}, \bbU_{A \cup B}\rangle = \delta \langle \bbU_A, \bbU_B \rangle 
= \langle Y_A, Y_B \rangle .
%\langle Y_A, Y_B\rangle = &
%\bigg\langle \sum_{S\in\binom{[n]}{\le d}} Q(S) \bbU_{S\cup A} ,
%\sum_{T\in\binom{[n]}{\le d}} Q(T) \bbU_{T\cup B} \bigg\rangle
%=  \sum_{S} Q(S) \bigg\langle \bbU_{S\cup A},
%\sum_{T} Q(T) \bbU_{T\cup B}
%\bigg\rangle \\
%= & \sum_{S,T} Q(S) Q(T) \langle \bbU_{S\cup A},
%\bbU_{T\cup B}
%\rangle  = \sum_{S,T} Q(S) Q(T) \langle \bbU_{S\cup A\cup B},
%\bbU_{T}
%\rangle
%\\
%= & \sum_{S} Q(S) \langle \bbU_{S\cup A\cup B},
%\sum_T Q(T) \bbU_{T}
%\rangle  =
%\sum_{S} Q(S) \langle \bbU_{S\cup A\cup B}, \delta \bbU_{\es}\rangle \\
%= & \delta \sum_{S} Q(S) \langle \bbU_{S},
%\bbU_{A\cup B}
%\rangle  = \delta \langle \sum_{S} Q(S) \bbU_{S},
%\bbU_{A\cup B}\rangle  = \delta \langle \delta \bbU_{\es},
%\bbU_{A\cup B}\rangle  \\
%= & \delta^2 \langle \bbU_{\es},
%\bbU_{A\cup B}\rangle   = \delta^2 \left\| \bbU_{A\cup B}\right\|^2.
%\tag*{\qedhere}
\end{align*}
\end{proof}

\subsection{Lasserre SDP for graph partitioning problems}

In light of \pref{thm:las-check}, to show good solutions for the Lasserre SDP for our problems of interest, we only need to show good solutions for the following SDPs.

\subsubsection{\balsep}
\label{sec:balsep}
The standard integer programming formulation of \balsep is shown in the left part of \pref{fig:balsep}. The $r$ round SDP relaxation $\Psi_1$ (shown in the right part of \pref{fig:balsep}) has a vector $\bbU_{S}$ for each subset $S \subseteq V$ with $|S| \le r$. In an integral solution, the intended value of $\bbU_{\{u\}}$ is $x_u \bbU_{\es}$ for some fixed unit vector $\bbU_{\es}$, and that of $\bbU_S$ is $\Bigl( \prod_{u \in S} x_u \Bigr) \bbU_{\es}$.

\begin{figure}[htbp]
  \caption{IP and SDP relaxations for \balsep. We can solve $\Psi_1$
    by first enumerating over all $\tau' \in \{1/n,2/n,\ldots,1\}\cap
    [\tau,1-\tau]$ and then choosing $\tau$ which minimizes the
    objective function. Note that the resulting relaxation is {\bf
      stronger} than usual Lasserre Hierarchy relaxation. }
 \label{fig:balsep}
\begin{tabular}{c|c}
\hline
{IP} & {SDP Relaxation $\Psi_1$}\\
\hline
\parbox{6cm}{\begin{align*}
\text{minimize} & ~~\sum_{(u, v)\in E}  (x_u - x_v)^2 \\
\text{s.t.} & ~~ \tau |V| \le \sum_{u\in V} x_u \le (1-\tau) |V| \\
& ~~  x_u \in \{0,1\} \quad \forall u \in V \ 
\end{align*}}&
\parbox{6cm}{\begin{align*}
\text{minimize} & \sum_{(u, v)\in E}  \norm{\bbU_{\{u\}} - \bbU_{\{v\}}}^2 \\
\text{s.t.} & \langle \bbU_{S_1}, \bbU_{S_2} \rangle \geq 0 \text{\ for all\ } S_1, S_2 \\
& \langle \bbU_{S_1}, \bbU_{S_2} \rangle = \langle \bbU_{S_3}, \bbU_{S_4} \rangle \text{\ for all\ } S_1 \cup S_2 = S_3 \cup S_4\\
& \norm{\bbU_{\emptyset}}^2 = 1 \\
& \sum_{v} \bbU_{\{v\}} = \tau' |V| \bbU_{\emptyset} \text{\ for some\ } \tau \leq \tau' \leq 1 - \tau
\end{align*}}\\
\hline
\end{tabular}
\end{figure}

\ynote{can we say something about this SDP and the Lasserre SDP? I.e. whether this SDP is stronger or the actually as strong as Lasserre...}

%Besides satisfying the general Lasserre conditions, we also require the cut is \emph{$\tau$-balanced}, i.e. $|V| \tau \le \sum_u x_u \le |V| (1-\tau)$. Equivalently:
%\vnote{This is technically two constraints and therefore doesn't exactly fit our SOS-IP, right? Should we have instead the constraint $\sum_{u < v} (x_u - x_v)^2 \ge \tau(1-\tau) |V|^2$ as this is the one in SOS form?}
%\aknote{Square both sides. Then it becomes an SOS constraint. Modified the constraint to reflect this.}
%In the Lasserre language, we write down the requirement as, for all $S \subseteq V$, $|S| %\leq r$, we have
%\begin{align*}
%\left( \sum_u x_u - \frac{|V|}{2}\right)^2 \le |V|^2 \left(\frac{1}{2} - \tau\right)^2.
%\tau |V| \cdot \|\bbU_S\|^2 \leq \sum_{v \in V} \langle \bbU_S, \bbU_{\{v\}} \rangle  \leq %(1 - \tau) |V|  \cdot \|\bbU_S\|^2.
%\end{align*}

\subsubsection{\scut}
\label{sec:scut}
%\aknote{Unfortunately this required quite a bit of reformulation. In the
%completeness
%proof, now we have to fix the value of $\tau$ a priori with a suitable lower
%bound. }
% For \scut, the goal of the Lasserre SDP is to minimize the sparsity (over all $\tau\in (0,1/2]$)
%\begin{eqnarray*}
%\frac{\sum_{(u, v)\in E}  \norm{\bbU_{\{u\}} - \bbU_{\{v\}}}^2 }{\frac{1}{2} \sum_{u, v \in V} \norm{\bbU_{\{u\}} - \bbU_{\{v\}}}^2} .
%\end{eqnarray*}
%\begin{align*}
%\frac{1}{|V|^2 \tau (1-\tau)} {\sum_{(u, v)\in E}  (x_u - x_v)^2},
%\end{align*}
%subject to $\sum_u x_u = \tau |V|$.
%, which can be expressed as: \vnote{Same comment as above.}
%\begin{align*}
%\left(\sum_u x_u - \tau |V|\right)^2 \le &  0.
%\end{align*}
The \scut\ problem asks to minimize the value of the quadratic integer program shown in the left part of \pref{fig:scut} over all $\tau \in \{1/n,2/n,\ldots,\lfloor n/2 \rfloor \}$. The corresponding SDP relaxation $\Psi_2$ is to minimize the value of the SDP shown in the right part of \pref{fig:scut} over all $\tau \in \{1/n,2/n,\ldots, \lfloor n/2 \rfloor\}$. 
%%%%
%\begin{remark}
\paragraph{Remark}
Prior to our paper, known %upper~\cite{ARV09,GS13,AGS13} and 
lower bounds~\cite{DKSV06,KaneM13} on the integrality gap of \scut\ problem used a {\bf weaker} relaxation, where 
the last two equality constraints in $\Psi_2$ of \pref{fig:scut}
are replaced by the following instead:
\[
\sum_{u<v} \|\bbU_{\{u\}} -
\bbU_{\{v\}}\|^2 = 1
\] with the objective function being simply $
\sum_{(u,v)\in E} \|\bbU_{\{u\}} - \bbU_{\{v\}}\|^2$.

%\end{remark}
%%%%
\begin{figure}[htbp]
  \caption{IP and SDP relaxations for \scut. We can solve $\Psi_2$ by
    first enumerating over all $\tau \in \{1/n,2/n,\ldots,(n-1)/n\}$
    and then choosing $\tau$ which minimizes the objective function. Note
    that the resulting relaxation is {\bf stronger} than usual
    Lasserre Hierarchy relaxation. }
 \label{fig:scut}
\begin{tabular}{c|c}
\hline
{IP} & {SDP Relaxation $\Psi_2$}\\
\hline
\parbox{6cm}{\begin{align*}
\text{minimize} & ~~\frac{1}{|V|^2 \tau (1-\tau)}\sum_{(u, v)\in E}  (x_u - x_v)^2 \\
\text{s.t.} & ~~  \sum_u x_u = \tau |V| \\
& ~~  x_u \in \{0,1\} \quad \forall u \in V \ 
\end{align*}}&
\parbox{6cm}{\begin{align*}
\text{minimize} & \sum_{(u, v)\in E}  \frac{1}{|V|^2 \tau (1 - \tau) }\norm{\bbU_{\{u\}} - \bbU_{\{v\}}}^2 \\
\text{s.t.} & \langle \bbU_{S_1}, \bbU_{S_2} \rangle \geq 0 \text{\ for all\ } S_1, S_2 \\
& \langle \bbU_{S_1}, \bbU_{S_2} \rangle = \langle \bbU_{S_3}, \bbU_{S_4} \rangle \text{\ for all\ } S_1 \cup S_2 = S_3 \cup S_4\\
& \norm{\bbU_{\emptyset}}^2 = 1 \\
& \sum_{v} \bbU_{\{v\}} = \tau |V| \bbU_{\emptyset}
\end{align*}}\\
\hline
\end{tabular}
\end{figure}

\subsection{Lasserre Gaps for \threexor from \cite{Sch08}}

We start by defining the \threexor problem.

\begin{definition}
An instance $\Phi$ of \threexor is a set of constraints $C_1, C_2, \cdots, C_m$ where each constraint $C_i$ is over $3$ distinct variables $x_{i_1}$, $x_{i_2}$, and $x_{i_3}$, and is of the form $x_{i_1} \oplus x_{i_2} \oplus x_{i_3} = b_i$ for some $b_i \in \{0, 1\}$. For each constraint $C_i$ and each partial assignment $\alpha$ that is valid on the variables $x_{i_1}$, $x_{i_2}$, and $x_{i_3}$, we use the notation $C_i(\alpha) = 1$ when $C_i$ is satisfied by $\alpha$ and $C_i(\alpha) = 0$ otherwise.

A random instance of \threexor is sampled by choosing each constraint $C_i$ uniform independently from the set of possible constraints.
\end{definition}

We will make use of the following fundamental result of Schoenebeck.
\begin{theorem}[\cite{Sch08}] \label{thm:Sch08}
For every large enough constant $\beta > 1$, there exists $\eta > 0$, such that with probability $1 - o(1)$, a random \threexor instance $\Phi$ over $m = \beta n$ constraints and $n$ variables cannot be refuted by the SDP relaxation obtained by $\eta n$ rounds of the Lasserre hierarchy, i.e. there are vectors $\bW_{(S, \alpha)}$ for all $|S| \leq \eta n$ and all $\alpha : S \rightarrow \{0, 1\}$, such that
\begin{enumerate}
\itemsep=0ex
\item[(i)] the value of the solution is perfect: $\sum_{i = 1}^{m} \sum_{\alpha : \{x_{i_1}, x_{i_2}, x_{i_3}\} \to \{0, 1\}, C_i(\alpha) = 1}   \norm{\bW_{(\{x_{i_1}, x_{i_2}, x_{i_3}\}, \alpha)}}^2 = m$;

\item[(ii)] $\langle \bW_{(S_1, \alpha_1)}, \bW_{(S_2, \alpha_2)}\rangle \geq 0$ for all $S_1, S_2, \alpha_1, \alpha_2$;

\item[(iii)] $\langle \bW_{(S_1, \alpha_1)}, \bW_{(S_2, \alpha_2)}\rangle = 0$ if $\alpha_1(S_1 \cap S_2) \neq \alpha_2(S_1 \cap S_2)$;

\item[(iv)] $\langle \bW_{(S_1, \alpha_1)}, \bW_{(S_2, \alpha_2)}\rangle = \langle \bW_{(S_3, \alpha_3)}, \bW_{(S_4, \alpha_4)}\rangle$ for all $S_1 \cup S_2 = S_3 \cup S_4$ and $\alpha_1 \circ \alpha_2 = \alpha_3 \circ \alpha_4$. Here, when $\alpha_1(S_1 \cap S_2) = \alpha_2(S_1 \cap S_2)$,  $\alpha_1 \circ \alpha_2$ is naturally defined as the mapping from $S_1 \cap S_2$ to $\{0, 1\}$ such that its restriction to $S_1$ equals $\alpha_1$ and its restriction to $S_2$ equals $\alpha_2$. We make similar definition for $\alpha_3 \circ \alpha_4$. 

\item[(v)] $\sum_{\alpha: S \rightarrow \{0, 1\}} \norm{\bW_{(S, \alpha)}}^2 = 1$ for all $S$.
\end{enumerate}
\end{theorem}
\noindent
Note that indeed we have for every $S$, $\sum_{\alpha: S \rightarrow \{0, 1\}} \bW_{(S, \alpha)} = \bW_{(\emptyset, \emptyset)}$. This is because $\norm{\bW_{(\emptyset, \emptyset)}}^2 = 1$ and
\begin{align*}
 \bigg\langle \Big( \sum_{\alpha: S \rightarrow \{0, 1\}} \bW_{(S, \alpha)} \Big), \bW_{(\emptyset, \emptyset)}\bigg\rangle = \sum_{\alpha: S \rightarrow \{0, 1\}}  \langle \bW_{(S, \alpha)} , \bW_{(\emptyset, \emptyset)}\rangle = \sum_{\alpha: S \rightarrow \{0, 1\}}  \norm{ \bW_{(S, \alpha)} }^2  = 1 .
\end{align*}
\begin{observation}
\label{obs:sch}
In the construction of \pref{thm:Sch08}, the vectors
$\bW$ satisfy the following property. For any constraint $C_i$ over set of variables $S_i$, the vectors corresponding to all satisfying partial assignments of $S_i$ sums up to $\bW_{\emptyset}$:
\[
\sum_{\alpha: S_i\to \{0,1\}\wedge C_i(\alpha) = 1} \bW_{(S_i,\alpha)} = \bW_{\emptyset}.
\]
\end{observation}
%\begin{proof}
%\aknote{Necessary?}
%\vnote{I'm okay with skipping..}
%\end{proof}
%\section{Lasserre Hierarchy Relaxation}

\section{Gaps for \balsep}
\label{sec:1}
In this section, we prove \pref{thm:informal-balsep}. We state the theorem in detail as follows.
\begin{theorem}\label{thm:bs}
Let $M$ be a large enough integral constant. For all $0.45 < \tau < 0.5$, and for infinitely many positive integer $N$'s, there is an $N$-vertex instance $\calH_\Phi$ for the $\tau$ vs. $(1 - \tau)$ \balsep problem, such that the optimal solution is at least $4(3\tau - \tau^3)/5 - O(1/M)$ times the best solution of the $\Omega(N)$-round Lasserre SDP relaxation. Moreover, the solution for Lasserre SDP relaxation is a fractional $(0.5 - O(1/M))$ vs. $(0.5 + O(1/M))$ balanced separator.
\end{theorem}

The rest of this section is dedicated to the proof of \pref{thm:bs}. In \pref{sec:bs-reduction}, we will describe how to get a \balsep instance from a \threexor instance. Then, we will show that when the \threexor instance is random, the corresponding \balsep instance is a desired gap instance. This is done by showing there is an SDP solution with good objective value (completeness part, \pref{lem:bs-completeness} in \pref{sec:bs-completeness}) while the instance in fact has not great integral solution (soundness part, \pref{lem:bs-soundness} in \pref{sec:bs-soundness}). The completeness part relies \pref{thm:Sch08} -- we use the \threexor vectors (which exist for random instances by the theorem) to construct \balsep vectors. In the soundness part, we first prove two pseudorandom structural properties exhibited in the random \threexor instances (\pref{lem:threexor-structure}), and then prove that any \threexor with these two properties leads to a \balsep instance with bad integral optimum by our construction. Finally, in \pref{sec:bs-constant-degree}, we slightly twist our gap instance in order to make its vertex degree bounded.

\subsection{Reduction} \label{sec:bs-reduction}

Given a \threexor instance $\Phi$ with $m = \beta n$ constraints and $n$ variables, we build a graph $\calH_\Phi = (\mathcal{V}_\Phi, \calE_\Phi)$ for \balsep as follows.

$\calH_\Phi$ consists of an almost bipartite graph $H_\Phi = (L_\Phi, R_\Phi, E_\Phi)$ (obtained by replacing each right vertex of a bipartite graph by a clique), an expander $Z_r$, and edges between $L_\Phi$ and $Z_r$.

The left side $L_\Phi$ of $H_\Phi$ contains $4m = 4 \beta n$ vertices, each corresponds to a pair of a constraint and a satisfying partial assignment for the constraint, i.e.
$$L_\Phi = \{(C_i, \alpha) | \alpha : \{x_{i_1}, x_{i_2}, x_{i_3}\} \rightarrow \{0, 1\}, C_i (\alpha) = 1 \} .$$
The right side $R_\Phi$ of $H_\Phi$ contains $2n$ cliques,  each is of size $M\beta$, and corresponds to one of the $2n$ literals, i.e.
$$R_\Phi = \cup_{j, \alpha : \{x_j\} \rightarrow \{0, 1\}} C_{(x_j, \alpha)} ,$$
where
$$ C_{(x_j, \alpha)} = \{(x_j, \alpha, t) |  1 \leq t \leq M \beta\} .$$
Call $(x_j, \alpha, 1)$ the \emph{representative vertex} of $C_{(x_j, \alpha)}$. Besides the clique edges, we connect a left vertex $(C_i, \alpha)$ and a right representative vertex $(x_j, \alpha', 1)$ if $x_j$ is accessed by $C_i$ and $\alpha'$ is consistent with $\alpha$, i.e.
$$E_\Phi = \{\text{clique edges}\} \cup \{\{(C_i, \alpha), (x_j, \alpha', 1)\} | x_j \in  \{x_{i_1}, x_{i_2}, x_{i_3}\}, \alpha(x_j) = \alpha'(x_j)\} .$$
Now we have finished the definition of $H_\Phi$. To get $\calH_\Phi$, we add an $O(M)$-regular expander $Z_r$ of size $m = \beta n$ and edge expansion $M$. (I.e. the degree of each vertex in $Z_r$ is $O(M)$, and each subset $T \subseteq Z_r$ ($|T| \leq |Z_r|/2$) has at least $|T| \cdot M$ edges connecting to $Z_r \setminus T$. For more discuss on the definitions and applications of expander graphs, please refer to, e.g.\ , \cite{HLW06}.) We connect each vertex in $L_\Phi$ to two different vertices in $Z_r$, so that each vertex in $Z_r$ has the same number of neighbors in $L_\Phi$ (this number should be $4\beta n \cdot 2 / (\beta n) = 8$). In other words, if we view each vertex in $L_\Phi$ as an undirected edge between its two neighbors in $Z_r$, the graph should be a regular graph.

\begin{figure}[h]
\includegraphics{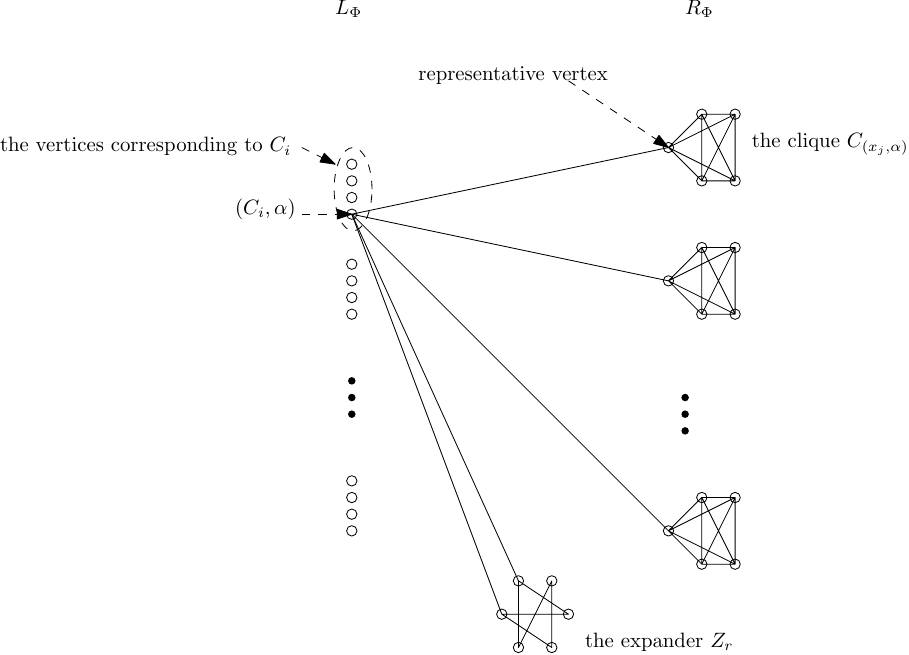}
\caption{The reduction for \balsep. Note that the incident edges are drawn for only one of the vertices in $L_\Phi$, while others can be drawn similarly.}  \label{fig:1}
\end{figure}

The whole construction is shown in \pref{fig:1}. Our construction is very similar to the one in \cite{AMS11}, but there are some technical differences. Instead of having cliques in $R_\Phi$, \cite{AMS11} has clusters of vertices with no edges connecting them. Also, in our construction, the vertices in $L_\Phi$ are connected to the representative vertices in $R_\Phi$ only, while in \cite{AMS11}, all the vertices in the right clusters could be connected to the left side. The most important difference is that in our way, the cliques are of constant size, while the clusters in \cite{AMS11} has superconstantly many vertices. This means that our reduction blows up the instance size only by a constant factor, therefore we are able to get linear round Lasserre gap. 

Observe that there are $|L_\Phi| + |R_\Phi| + |Z_r| = 4m + 2M m + m = (2M + 5)m$ vertices in $\calH_\Phi$.

In the following two subsections, we will prove the completeness lemma (\pref{lem:bs-completeness}, which states that there is an SDP solution with a good objective value) and the soundness lemma (\pref{lem:bs-soundness}, which states that every integral solution has a bad objective value). Combining the two lemmas, we prove our main integrality gap theorem for \balsep as follows.
\begin{proof}[of \pref{thm:bs} from \pref{lem:bs-completeness} and \pref{lem:bs-soundness}]
Let $\beta, M$ be large enough constants. Let $\Phi$ be a random \threexor instance over $m = \beta n$ constraints and $n$ variables. For all $0.45 < \tau < 0.5$, we will show that the optimal solution for  $\calH_\Phi$  is at least $4(3\tau - \tau^3)/5 - O(1/\sqrt{\beta} + 1/M)$ times the best solution of the $\Omega(N)$-round Lasserre SDP relaxation; and the solution for Lasserre SDP relaxation is a fractional $(0.5 - O(1/M))$ vs. $(0.5 + O(1/M))$ balanced separator. One may choose $\beta = M^2$ to get the statement in the theorem.

By \pref{thm:Sch08} we know that, with probability $1 - o(1)$, $\Phi$ admits a perfect solution for $\Omega(n)$-round Lasserre SDP relaxation. Therefore, by \pref{lem:bs-completeness}, with probability $1 - o(1)$, $\Omega(n)$-round SDP relaxation $\Psi_1$ with parameter $\tau = 0.5 - O(1/M)$ for the \balsep instance $\calH_\Phi$ has a solution of value $5m$.  On the other hand, by \pref{lem:bs-soundness}, with probability $1 - o(1)$, for $\tau > 1/3$, every $\tau$ vs. $(1-\tau)$ balanced separator has at least  $4m (3 \tau  - \tau^3 - O(1/ \sqrt{\beta}) - O(1/M) )$ edges in the cut.

Therefore, with probability $1 - o(1)$, when $\tau > 1/3$, the ratio between the optimal integral solution (to $\calH_\Phi$) and the optimal $\Omega(n)$-round $\Psi_1$ solution is at least $4(3\tau - \tau^3)/5 - O(1/\sqrt{\beta} + 1/M)$. This ratio is greater than $1.007$ when $\tau > 0.45$ and $\beta$ and $M$ are large enough. By our observation in \pref{sec:balsep}, this gap also holds for the Lasserre SDP relaxation.
\end{proof}

Let $\Delta$ be the maximum number of occurrences of any variable in $\Phi$. By our construction, the graph has degree $\Theta(M + \Delta)$. When $\beta = O(1)$, we have $\Delta = \Theta(\log n / \log \log n)$ with probability $1 - o(1)$ (see, e.g.\ \cite{Gon81}). This means that our graph does not have the desired constant-degree property. However, since there are few edges incident to vertices with superconstant degree, we can simply remove all these edges to get a constant-degree graph, while the completeness and soundness are still preserved. We will discuss this in more details in \pref{sec:bs-constant-degree}.

\subsection{Completeness : good SDP solution} \label{sec:bs-completeness}
%\newcommand{\deferapx}[1]{The proof of the following is given in
%Appendix~\ref{#1}.}
%\deferapx{apx:1}
%\newcommand{\bsclemma}
\ 

\begin{lemma}[Completeness]\label{lem:bs-completeness}
If the \threexor instance $\Phi$ admits perfect solution for $r$-round Lasserre SDP relaxation, % and satisfies the first condition in \pref{lem:threexor-structure}, 
then the $r/3$-round SDP relaxation $\Psi_1$ (in \pref{fig:balsep}) with parameter $\tau = 0.5 - O(1/M)$ for the \balsep instance $\calH_\Phi$ has a solution of value $5m$.
\end{lemma}

\begin{proof}
We define a set of vectors (i.e. a solution to $\Psi_1$) using the vectors given in \pref{thm:Sch08}, as follows.

For each set $S \subseteq L_\Phi \cup R_\Phi \cup Z_r$ with $|S| \leq r /3$, we define the vector $\bbU_S$ as follows. If $S \cap Z_r \neq \emptyset$, let $\bbU_S = \bm{0}$. If $S \cap Z_r = \emptyset$, suppose that $S \cap L_\Phi$ contains
\begin{align*}
(C_{i_1}, \alpha_1), (C_{i_2}, \alpha_2), \cdots, (C_{i_{r_1}}, \alpha_{r_1}),
\end{align*}
$S \cap R_\Phi$ contains
\begin{align*}
(x_{j_1}, \alpha_{1}', t_{1}), (x_{j_2}, \alpha_{2}', t_{2}), \cdots, (x_{j_{r_2}}, \alpha_{{r_2}}', t_{r_2}),
\end{align*}
we have $r_1 + r_2 = |S|$. Let $S'$ be the set of variables accessed by $C_{i_1}, \cdots C_{i_2}$ together with $x_{j_1}, \cdots, x_{j_{r_2}}$. Note that $|S'| \leq 3 r_1 + r_2 \leq 3 |S| \leq r$. If there is no contradiction among the partial assignments $\alpha_i$'s and $\alpha_i'$'s (i.e. there are not two of them assigning the same variable to different values), we can define
\begin{align*}
\alpha = \alpha_1 \circ \alpha_2 \circ \cdots \alpha_{r_1} \circ \alpha_1' \circ \alpha_2' \circ \cdots \circ \alpha_{r_2}' .
\end{align*}
and let $\bbU_S = \bW_{(S', \alpha)}$, otherwise we let $\bbU_S = \bm{0}$.

We first check that the first 3 constraints in relaxation $\Psi_1$ are satisfied.
\begin{itemize}
\item For two sets $S_1, S_2$, either at least one of the vectors $\bbU_{S_1}, \bbU_{S_2}$ is $\bm{0}$ (therefore their inner-product is $0$), or $\bbU_{S_1} = \bW_{S_1', \alpha_1}, \bbU_{S_2} = \bW_{S_2', \alpha_2}$ for some $S_1', S_2', \alpha_1, \alpha_2$ and $\langle \bbU_{S_1}, \bbU_{S_2} \rangle = \langle \bW_{S_1', \alpha_1}, \bW_{S_2', \alpha_2} \rangle \geq 0$.

\item For any $S_1, S_2, S_3, S_4$ such that $S_1 \cup S_2 = S_3 \cup S_4$, either the set of partial assignments in $S_1 \cup S_2 = S_3 \cup S_4$ are consistent with each other, in which case we have $\bbU_{S_1 \cup S_2} = \bbU_{S_3 \cup S_4} = \bW_{S, \alpha}$ where $S$ is the union of all the variables included in $S_1 \cup S_2$ and $\alpha$ is the concatenation of the partial assignments in $S_1 \cup S_2$; or we have $\bbU_{S_1 \cup S_2} = \bbU_{S_3 \cup S_4} = \bm{0}$.

\item  $\norm{\bU_{\emptyset}}^2 = \norm{\bW_{(\emptyset, \emptyset)}}^2 = 1$.
\end{itemize}
\let\es=\emptyset
Now we check that the balance condition (the last constraint in relaxation $\Psi_1$) is satisfied.
We will prove that
\[
\sum_v \bbU_{\{v\}} = (M+1) m   \bbU_{\es}.
\]
Since there are $(2M + 5)m$ vertices in $\calH_\Phi$, this shows that the solution is feasible for $\Phi_1$ with $\tau = 0.5 - O(1/M)$.
Using \pref{obs:sch}, we see that $\sum_{(C_i, \alpha) \in L_\Phi} \bbU_{\{(C_i, \alpha)\}}
= \sum_{C_i} \bbU_{\es} = m \bbU_{\es} $.
Similarly
\begin{align*}
\sum_{(x_j, \alpha, t) \in R_\Phi}\bbU_{\{(x_j, \alpha, t)\}} = &
\sum_{j = 1}^{n} \sum_{\alpha : \{x_j\} \rightarrow \{0, 1\}} \sum_{t=1}^{\beta M}
 \bbU_{\{(x_j, \alpha, t)\}}
 = \beta M\cdot  \sum_{j = 1}^{n} \sum_{\alpha : \{x_j\} \rightarrow \{0, 1\}}
  \bbU_{\{(x_j, \alpha, 1)\}}  \\
  =& \beta M n \cdot \bbU_{\es} = M m\bbU_{\es} .
\end{align*}
Thus
\[
\sum_{v\in V} \bbU_{\{v\}} =
\sum_{v \in L_\Phi \cup R_\Phi \cup Z_r} \bbU_{\{v\}}
=
\sum_{(C_i, \alpha) \in L_\Phi} \bbU_{\{(C_i, \alpha)\}}+\sum_{(x_j, \alpha, t) \in R_\Phi}\bbU_{\{(x_j, \alpha, t)\}}
= (M+1) m  \bbU_{\es}.
\]
%For all $|S| \leq r$, we have
%\begin{align*}
%& \sum_{v \in L_\Psi \cup R_\Phi \cup Z_r} \langle \bbU_S, \bbU_{\{v\}}\rangle = \sum_{(C_i, \alpha) \in L_\Phi} \langle \bbU_S, \bbU_{\{(C_i, \alpha)\}}\rangle + \sum_{(x_j, \alpha, t) \in R_\Phi} \langle \bbU_S, \bbU_{\{(x_j, \alpha, t)\}} \rangle\\
%=&  \sum_{(C_i, \alpha) \in L_\Phi} \langle \bbU_S, \bbU_{\{(C_i, \alpha)\}}\rangle + \sum_{j = 1}^{n} \sum_{\alpha : \{x_j\} \rightarrow \{0, 1\}} \sum_{t=1}^{\beta M}  \langle \bbU_S, \bbU_{\{(x_j, \alpha, t)\}} \rangle\\
%=&  \sum_{(C_i, \alpha) \in L_\Phi} \langle \bbU_S, \bbU_{\{(C_i, \alpha)\}}\rangle + \beta M\cdot \sum_{j = 1}^{n} \sum_{\alpha : \{x_j\} \rightarrow \{0, 1\}}  \langle \bbU_S, \bbU_{\{(x_j, \alpha, t)\}} \rangle\\
%=&  \sum_{(C_i, \alpha) \in L_\Phi} \langle \bbU_S, \bbU_{\{(C_i, \alpha)\}}\rangle + \beta M\cdot \sum_{j = 1}^{n}  \norm{\bbU_S}^2\\
% \in&  [Mm \cdot \norm{\bbU_{S}}^2, (M + 4)m \cdot \norm{\bbU_{S}}^2] .
%\end{align*}

Now, we calculate the value of the solution
\begin{align*}
&\sum_{(u,v) \in \calE_\Phi} \norm{\bbU_{\{u\}} - \bbU_{\{v\}}}^2\\
 =& \sum_{i = 1}^{m} \sum_{\alpha : \{x_{i_1}, x_{i_2}, x_{i_3}\} \rightarrow \{0, 1\}, C_i(\alpha)= 1} \sum_{z = 1}^{3} \norm{\bbU_{\{(C_i, \alpha)\}} - \bbU_{\{(x_{i_z}, \alpha_{|\{x_{i_z}\}}, 1)\}}}^2\\
 & \qquad + \sum_{i = 1}^{m} \sum_{\alpha : \{x_{i_1}, x_{i_2}, x_{i_3}\} \rightarrow \{0, 1\}, C_i(\alpha)= 1} \ \sum_{v \in Z_r : ((C_i, \alpha), v) \in \calE_\Phi} \norm{\bbU_{\{(C_i, \alpha)\}} - \bbU_{\{v\}}}^2  \\
 & \qquad + \sum_{j = 1}^{n} \sum_{\alpha : \{x_j\} \rightarrow \{0, 1\}} \sum_{z_1, z_2 \in [M\beta]} \norm{\bbU_{\{(x_j, \alpha, z_1)\}} - \bbU_{\{(x_j, \alpha, z_2)\}}}^2 + \sum_{v_1, v_2 \in Z_r} \norm{\bbU_{\{v_1\}} - \bbU_{\{v_2\}}}^2\\
 =& \sum_{i = 1}^{m} \sum_{\alpha : \{x_{i_1}, x_{i_2}, x_{i_3}\} \rightarrow \{0, 1\}, C_i(\alpha)= 1} \left( \sum_{z = 1}^{3} \norm{\bbU_{\{(C_i, \alpha)\}} - \bbU_{\{(x_{i_z}, \alpha_{|\{x_{i_z}\}}, 1)\}}}^2 + 2 \norm{\bbU_{\{(C_i, \alpha)\}}}^2 \right)\\
 =& \sum_{i = 1}^{m} \sum_{\alpha : \{x_{i_1}, x_{i_2}, x_{i_3}\} \rightarrow \{0, 1\}, C_i(\alpha)= 1} \left( \sum_{z = 1}^{3} \norm{\bW_{(\{x_{i_1}, x_{i_2}, x_{i_3}\}, \alpha)} - \bW_{(\{x_{i_z}\}, \alpha_{|\{x_{i_z}\}})}}^2 + 2 \norm{\bW_{(\{x_{i_1}, x_{i_2}, x_{i_3}\}, \alpha)}}^2 \right)\\
 =& \sum_{i = 1}^{m} \sum_{\alpha : \{x_{i_1}, x_{i_2}, x_{i_3}\} \rightarrow \{0, 1\}, C_i(\alpha)= 1} \left( \sum_{z = 1}^{3} \left( \norm{\bW_{(\{x_{i_z}\}, \alpha_{|\{x_{i_z}\}})}}^2 -\norm{\bW_{(\{x_{i_1}, x_{i_2}, x_{i_3}\}, \alpha)}}^2\right) + 2 \norm{\bW_{(\{x_{i_1}, x_{i_2}, x_{i_3}\}, \alpha)}}^2 \right)\\
 =& \sum_{i = 1}^{m} \sum_{\alpha : \{x_{i_1}, x_{i_2}, x_{i_3}\} \rightarrow \{0, 1\}, C_i(\alpha)= 1}  \sum_{z = 1}^{3}  \norm{\bW_{(\{x_{i_z}\}, \alpha_{|\{x_{i_z}\}})}}^2  - \sum_{i = 1}^{m} \sum_{\alpha : \{x_{i_1}, x_{i_2}, x_{i_3}\} \rightarrow \{0, 1\}, C_i(\alpha)= 1}  \norm{\bW_{(\{x_{i_1}, x_{i_2}, x_{i_3}\}, \alpha)}}^2 \\
 =& \sum_{i = 1}^{m} \sum_{z=1}^{3} 2 \left(\norm{\bW_{\{x_{i_z}\}, \{x_{i_z} \rightarrow 0\}}}^2 + \norm{\bW_{\{x_{i_z}\}, \{x_{i_z} \rightarrow 1\}}}^2 \right) - m\\
 =& 6m - m = 5m .
\end{align*}
\end{proof}

\subsection{Soundness : bound for integral solutions} \label{sec:bs-soundness}

 Let $\calL = \{(x_j, \alpha) | \alpha : \{x_j\} \rightarrow \{0, 1\}\}$ be the set of $2n$ literals. For each literal $(x_j, \alpha) \in \calL$, let $\deg((x_j, \alpha))$ be the number of left vertices that connect to the literal's representative vertex $(x_j, \alpha, 1)$. For a set of literals $\calL' \subseteq \calL$, let $\deg(\calL') = \sum_{(x_j, \alpha) \in \calL'} \deg((x_j, \alpha))$. Also, given a subset $\calL' \subseteq \calL$, for left vertex $(C_i, \alpha)$, say $(C_i, \alpha)$ is \emph{contained} in $\calL'$ if all the three literals corresponding to the three neighbors of $(C_i, \alpha)$ in $H_\Phi$ are contained in $\calL'$, i.e. $\{(x_{i_1}, \alpha_{|x_{i_1}}), (x_{i_2}, \alpha_{|x_{i_2}}), (x_{i_3}, \alpha_{|x_{i_3}})\} \subseteq \calL'$.

We first prove the following lemma regarding the structure of $\calH_\Phi$, defined by a random \threexor instance $\Phi$.
\begin{lemma}\label{lem:threexor-structure}
%\deferapx{apx:1}
%\newcommand{\threexorlem}
{
Over the choice of random \threexor instance $\Phi$, with probability $1 - o(1)$, the following statements hold.
\begin{itemize}
\item For each $\calL' \subseteq \calL$, $|\calL'| \geq n / 3$, we have $\deg(\calL') \geq 6m\cdot|\calL'|/n (1- 20/\sqrt{\beta})$.
\item For each $\calL' \subseteq \calL$, $|\calL'| \geq n/3$, the number of left vertices in $L_\Phi$ contained in $\calL'$ is at most $m \cdot |\calL'|^3/(2n^3) \cdot (1 + 100/\sqrt{\beta})$.
\end{itemize}
}
\end{lemma}

\begin{proof}
Fix a literal $(x_j, \alpha)$, a random constraint $C_i$ accesses $x_j$ with probability $3/n$. Once $C_i$ accesses $x_j$, there are $2$ vertices out of the $4$ left vertices corresponding to $C_i$ adjacent to $(x_j, \alpha)$. Therefore, in expectation, there are $6/n$ edges from the left vertices corresponding to $C_i$ to $(x_j, \alpha)$. By linearity of expectation, for fixed $\calL' \subseteq \calL$, there are $6 |\calL'|/n$ edges from the left vertices corresponding to a random constraint $C_i$ to $\calL'$ in expectation.

Now for each $C_i$, let the random variable $X_i$ be the number of representative vertices in $\calL'$ that is connected to left vertices corresponding to $C_i$. By definition we have $\deg(\calL') = \sum_{i=1}^{m} X_i$. Since each left vertex corresponding to $C_i$ has $3$ neighbors on the right side, and there are $4$ of such left vertices, we know that $X_i \in [0, 12]$. In the previous paragraph we have concluded that $\E[X_i] = 6|\calL'|/n$ for all $i = 1, 2, \dots, m$. It is also easy to see that $X_1, X_2, \dots, X_m$ are independent random variables.

Now assuming that $|\calL'| \geq n/3$, we use Hoeffding's inequality for the random variables $X_1, X_2, \dots, X_m$, and get
\begin{align*}
 &\Pr[\deg(\calL') < 6m\cdot |\calL'|/n(1 - 20/\sqrt{\beta})] 
 = \Pr\left[ \sum_{i=1}^{n} X_i < 6m\cdot |\calL'|/n(1 - 20/\sqrt{\beta})\right]\\
  \leq  &\exp\left(-\frac{2 \cdot \left(\frac{20}{\sqrt{\beta}} \cdot 6m \cdot \frac{|\calL'|}{n}\right)^2}{m \cdot 12^2}\right) 
 = \exp \left(-200 \cdot \left(\frac{|\calL'|}{n}\right)^2 \cdot n\right)
  \leq   \exp\left(-22 n\right)
   \leq   2^{-4n} .
\end{align*}
Since there are at most $2^{2n}$ such $\calL'$'s, by a union bound, with probability at least $1-2^{-2n}$, the first statement holds.

For the second statement, fix an $\calL' \subseteq \calL$, let $a_0, a_1, a_2$ be the number of variables that have $0, 1, 2$ corresponding literals in $\calL'$, respectively. Note that $a_0 + a_1 + a_2 = n$ and $a_1 + 2a_2 = |\calL'|$ Now, for a random constraint $C_i$, we are interested in the expected number of the four corresponding left vertices $(C_i, \alpha)$ that are contained in $\calL'$. Note that once $C_i$ accesses a variable that corresponds to $a_0$, none of the four corresponding left vertices are contained in $\calL'$. Now let us condition on the case that, out of the $3$ variables accessed by $C_i$, $t$ variables have two literals in $\calL'$ and the other $(3 - t)$ variables have one literal in $\calL'$. Observe that in expectation (which is over the random choice of $C_i$ while conditioned on $t$), there are $2^{t-1}$ left vertices corresponding to $C_i$ contained in $\calL'$.

%There are $2^{t}$ partial assignments for the $3$ variables that are contained in $\calL'$, and each of the partial assignment satisfies $C_i$ with probability $1/2$. Therefore, conditioned on $t$ is fixed, the expectation of the left vertices corresponding to $C_i$ that are contained in $\calL'$ is $2^{t-1}$.

In all, the expected number of the left vertices corresponding to $C_i$ that are contained in $\calL'$ is
\begin{align*}
\sum_{t = 0}^{3} \frac{{a_1 \choose 3- t}{a_2 \choose t}}{{n \choose 3}} \cdot 2^{t-1} & < \left(1 + \frac{10}{n}\right)  \sum_{t = 0}^{3} {3 \choose t} (a_1/n)^{3-t} (a_2 / n)^t\cdot  2^{t-1} & \text{(for $n > 3$)}\\
&= \left(1 + \frac{10}{n}\right) (a_1 + 2a_2)^3 / (2n^3) = \left(1 + \frac{10}{n}\right) \cdot |\calL'|^3 / (2n^3) .
\end{align*}
For each $C_i$, let the random variable $X_i$ be the number of left vertices corresponding to $C_i$ that are contained in $\calL'$. By the discuss above, we know that $\E [X_i] < \left(1 + \frac{10}{n}\right) \cdot |\calL'|^3 / (2n^3)$. Now we are interested in the probability that the total number of left vertices contained in $\calL'$ (i.e. $\sum_{i=1}^{m} X_i$) is big. Since $X_i$'s are always bounded by $[0, 4]$,
by standard Chernoff bound, we have
\begin{align*}
&\Pr\left[\sum_{i=1}^{m} X_i >   m\cdot |\calL'|^3 / (2n^3) \cdot (1 +100/\sqrt{\beta})\right] \\
= &\Pr\left[\sum_{i=1}^{m} X_i >   m\cdot \left(1 + \frac{10}{n}\right) \cdot  |\calL'|^3 / (2n^3)  \cdot \frac{1+100/\sqrt{\beta}}{1 + 10/n}\right] \\
= & \Pr\left[\sum_{i=1}^{m} X_i >   m\cdot \left(1 + \frac{10}{n}\right) \cdot  |\calL'|^3 / (2n^3)  \cdot \left(1 +\frac{100/\sqrt{\beta} - 10/n}{1 + 10/n}\right)\right]\\
\leq & \exp\left(-\frac{1}{4} \cdot m \cdot \left(1 + \frac{10}{n}\right) \cdot  |\calL'|^3 / (2n^3)  \cdot \frac{(100/\sqrt{\beta} - 10/n)^2}{3 (1 + 10/n)^2}\right) & (\text{for large enough $\beta$})\\
%\leq &\exp\left(-m\cdot |\calL'|^3 / (2n^3)/4 \cdot ((100 /\sqrt{\beta} - 10/n) \cdot n / (n + 10) )^2 / 2\right) \\
\leq &\exp\left(-\frac{1}{4} \cdot m\cdot |\calL'|^3 / (2n^3) \cdot \frac{(80/\sqrt{\beta} )^2 }{3}\right) & \text{(for $n \gg \sqrt{\beta} \gg 1$)} \\
= &\exp\left(-\beta n \cdot \frac{|\calL'|^3}{n^3} \cdot \frac{1}{\beta} \cdot \frac{800}{3}\right)\\
\leq &\exp\left(- n\ \cdot \frac{800}{3^4}\right)  & \text{(since $|\calL'| \geq n/3$)}\\
\leq& 2^{-4n} .
\end{align*}
Since there are at most $2^{2n}$ such $\calL'$'s, by a union bound, with probability at least $1 - 2^{-2n}$, the second statement holds.
\end{proof}

Now, we are ready to prove the soundness lemma.
\begin{lemma}[Soundness]\label{lem:bs-soundness}
For $\tau > 1/3$, with probability $1 - o(1)$, the $\tau$ vs. $(1-\tau)$ balanced separator has at least  $4m (3 \tau  - \tau^3 - O(1/ \sqrt{\beta}) - O(1/M) )$ edges in the cut.
\end{lemma}
\begin{proof}
We are going to prove that, once the two conditions in \pref{lem:threexor-structure} hold, we have the desired upper bound for $\tau$ vs. $(1-\tau)$ balanced separator. Let us assume that there is a balanced separator $(A', B')$ such that $\edges(A', B') \leq 4m (3 \tau  - \tau^3) \leq 12 m $, we will show that $\edges(A', B') \geq 4m (3 \tau  - \tau^3 - O(1/ \sqrt{\beta}) - O(1/M) )$.

Based on $(A', B')$ we build another cut $(A, B)$ such that $A \cap Z_r = A' \cap Z_r$ and $A \cap R_\Phi = A' \cap R_\Phi$. For each left vertex in $L_\Phi$, it has $5$ edges going to $Z_r$ and $R_\Phi$. We assign the vertex to $A$ if it has less than $3$ edges going to $B' \cap (Z_r \cup R_\Phi)$, and assign it to $B$ otherwise. Note that $\edges(A, B) \leq \edges(A', B')$, therefore we only need to show that $\edges(A, B) \geq m (12 \tau  - \tau^3 - O(1/ \sqrt{\beta}) - O(1/M) )$. Since $L_\Phi$ contains only $O(1/M)$ fraction of the total vertices, $(A, B)$ is still $(\tau - O(1/M))$ vs.  $(1 - \tau + O(1/M))$ balanced.

Since $\edges(A, B) \leq 12m$, for large enough constant $M$, we have the following two statements.
\begin{itemize}
\item[1)] One of  $A \cap Z_r$ and $B \cap Z_r$ has at most $100/M \cdot |Z_r| = 100m/M$ vertices.
\item[2)] Let $\calC_{\bad} = \{(x_j, \alpha) : \text{the clique $C_{(x_j, \alpha)}$ is broken by $(A, B)$}\}$, then $|\calC_{\bad}| \leq 20n/M$.
\end{itemize}
If 1) does not hold, then we see there are at least $(100/M)\cdot |Z_r| \cdot M = 100m$ edges in $Z_r$ cut by $(A, B)$, by the expansion property.  If 2) does not hold, for each clique $C_{(x_j, \alpha)}$ that is broken by $(A, B)$, at least $(\beta M - 1)$ edges of the clique are in the cut. In all, there are at least $(\beta M - 1) \cdot 20n / M > 12 \beta n = 12m$ edges in the cut.

Now, by 1), assume w.l.o.g. that $A\cap Z_r$ is the smaller side -- having at most $100/M \cdot |Z_r|$ vertices, and let $\calL' $ be the set of literals $(x_j, \alpha)$ such that its representative vertex $(x_j, \alpha, 1)$ is in $A$. %To get an upperbound for $|\calL'|$, note that $A$ contains all the cliques corresponding to $\calL' \setminus \calC_{\bad}$, therefore $|A| \geq (|\calL'| - \calC_\bad|) M\beta \geq |\calL'| \cdot M \beta - 3 m$ (by 2) ). On the other hand, since $(A, B)$ is a balanced separator, we know that $ |A| \leq (1 - \tau + O(1/M)) (2M + O(1))m$. In all, we have $|\calL'| \cdot M\beta - 3 m \leq (1 - \tau + O(1/M))(2M + O(1))m$, i.e. $|\calL'|  \leq (1- \tau)(2 + O(1/M)) n$.

To get a lower bound for $|\calL'|$, note that
\begin{eqnarray}
|A| \leq (|\calL'| + |\calC_\bad|) \cdot M \beta + |Z_r| + |L_\Phi| = |\calL'| \cdot M \beta + O(1) m . \label{eqn:1}
\end{eqnarray}
Also, since $(A, B)$ is a balanced separator, we have $|A| \geq (\tau - O(1/M)) \cdot 2M m$. Hence, by \eqref{eqn:1}, we have $|\calL'| \geq (\tau - O(1/M)) \cdot 2n$.

Let $L_\bad \subseteq L_\Phi$ be the set of left vertices such that at least one of the two neighbors in $Z_r$ falls into $A \cap Z_r$. By the regularity of the graph where $Z_r$ is the set of vertices and $L_\Phi$ is the set of edges, we know that $|L_\bad| \leq 8 \cdot 100 / M \cdot |Z_r| \leq O(m / M) $.

Now let us get a lower bound on $\edges(A, B)$. First, we have $\edges(A, B) \geq \edges(A \setminus L_\bad, B\setminus L_\bad)$. Let $L_\Phi' = L_\Phi \setminus L_\bad$, we have
\begin{align*}
&~ \edges(A \setminus L_\bad, B\setminus L_\bad) \\
=  &~ \edges(A \cap (L_\Phi' \cup R_\Phi \cup Z_r), B\cap(L_\Phi' \cup R_\Phi \cup Z_r))\\
 \geq &~ \edges(A \cap L_\Phi', B \cap Z_r) + \edges(A \cap R_\Phi, B \cap L_\Phi')\\
 = &~ \edges(A \cap L_\Phi', B \cap Z_r) + \edges(A \cap R_\Phi,  L_\Phi') - \edges(A \cap R_\Phi, A \cap L_\Phi')\\
 \geq&~  \edges(A \cap L_\Phi', B \cap Z_r) + \edges(A \cap R_\Phi,  L_\Phi) - |L_\bad| \cdot 3 - \edges(A \cap R_\Phi, A \cap L_\Phi').
\end{align*}

Consider a left vertex $(C_i, \alpha) \in L_\Phi'$. We claim that it is contained in $\calL'$ if and only if $(C_i, \alpha) \in A$. This is because if it is contained in $\calL'$, then we have $(C_i, \alpha) \in A$ because  $3$ out of $5$ edges incident to $(C_i, \alpha)$ go to $A$ side (the three variable representative vertices). If $(C_i, \alpha)$ is not contained in $\calL'$, we have at least $3$ out of the $5$ edges going to $B$  side (the two edges to $B \cap Z_r$ and at least one of the variable representative vertices), and therefore we have $(C_i, \alpha) \in B$.  By this claim, we know the following two facts.
\begin{itemize}
\item $|A \cap L_\Phi'|$ is small. Since $\tau > 1/3$, we have $|\calL'| \geq (2/3 - O(1/M)) n > n/3$, and by the second property of \pref{lem:threexor-structure}, we have $|A \cap L_\Phi'| \leq m \cdot |\calL'|^3/(2n^3) \cdot (1 + 100/\sqrt{\beta})$.
\item We have $\edges(A \cap L_\Phi', B \cap Z_r) = 2 |A \cap L_\Phi'|$ and $\edges(A \cap L_\Phi', A \cap R_\Phi) = 3|A \cap L_\Phi'|$.
\end{itemize}

For $\edges(A \cap R_\Phi, L_\Phi)$, we know that this is exactly $\deg(\calL')$. Again, since $\tau > 1/3$, by the first property of \pref{lem:threexor-structure}, we know this value is lower-bounded by $6m \cdot |\calL'| / n (1 - 20/\sqrt{\beta})$.

In all, we have
\begin{align*}
\edges(A, B) &\geq \edges(A \cap L_\Phi', B \cap Z_r) + \edges(A \cap R_\Phi,  L_\Phi) - |L_\bad| \cdot 3 - \edges(A \cap R_\Phi, A \cap L_\Phi')\\
& = 2|A \cap L_\Phi'| + \deg(\calL') - |L_\bad| \cdot 3 - 3|A \cap L_\Phi'|\\
& \geq \deg(\calL') - |A \cap L_\Phi'| - O(m/M)\\
& \geq 6m \cdot |\calL'| / n (1 - 20/\sqrt{\beta}) -  m \cdot |\calL'|^3/(2n^3) \cdot (1 + 100/\sqrt{\beta}) - O(m/M) \\
& = m \left(12 \gamma  - 4\gamma^3 - (240 \gamma + 400\gamma^3)/ \sqrt{\beta} - O(1/M) \right) \qquad\qquad (\text{let $\gamma = |\calL'| / (2n)$})\\
& \geq 4m \left(3 \tau  - \tau^3 - O(1/ \sqrt{\beta}) - O(1/M) \right) .
\end{align*}
The last step follows because (i) $3 \gamma  - \gamma^3$ monotonically increases when $\gamma \in [0, 1]$, and (ii) $\gamma \geq (\tau - O(1/M))$.
\end{proof}

\subsection{Constant-degree integrality gap instance}\label{sec:bs-constant-degree}

In this subsection, we slightly modify the graph $\calH_\Phi$ obtained in the previous subsections to get an integrality gap instance with constant degree. 

Observe that in $\calH_\Phi$, when $M$ and $\beta$ are constants, the only vertices whose degree might be superconstant are the representative vertices in $R_{\Phi}$. Now consider the edges connecting vertices in $L_\Phi$ and representative vertices: there are $12m$ of them, each of them corresponds to a combination of constraint $C_i$, satisfying assignment $\alpha$, and one of the variables in the constraint. Let $E_b$ be the set of these edges.

For two edges $e_1, e_2 \in E_b$, let the random variable $Y_{\{e_1, e_2\}} = 1$ if they share the same representative vertex, and let $Y_{\{e_1, e_2\}} = 0$ otherwise. Finally let $Y = \sum_{e_1, e_2 \in E_b} Y_{\{e_1, e_2\}}$. By the simple second moment method, we know that with probability $1 - o(1)$, we have $Y \leq \frac{1000m^2}{n} = 1000 \beta^2 n$.

For every edge $e \in E_b$, if $\sum_{e' \in E_b \setminus \{e\}} Y_{\{e, e'\}} > \beta M$, we remove $e$ from the graph. In this way, we get a new graph, namely $\calH_\Phi'$. We claim the following properties about $\calH_\Phi'$.
\begin{enumerate}
\item The maximum degree of $\calH_\Phi'$ is $O(\beta M)$. This is because the maximum degree of vertices other than representative vertices in $\calH_\Phi$ is $O(\beta M)$, and after the edge removal process described above, the representative vertices have degree $O(\beta M)$.

\item The number of edges removed is at most $2Y / (\beta M)$, and therefore $2000 m/M$ with probability $1 - o(1)$. This is because whenever an edge is removed, we charge $\beta M$ to $Y$. Since each edge in $Y$ can be charge at most twice, there are at most $2Y / (\beta M)$ edges to be removed.

\item The SDP solution in \pref{lem:bs-completeness} is still feasible and has objective value at most $5m$ (since we removed edges) with probability $1 - o(1)$. 

\item The soundness lemma \pref{lem:bs-soundness} still holds since we removed only $O(m/M)$ edges. 
\end{enumerate}

Therefore, we claim that $\calH_\Phi'$ is an integrality gap instance for \pref{thm:bs} with constant degree. 

\section{Gaps for \scut}
\label{sec:2}

In this section, we provide the full analysis of the gap instance for \scut. We first describe our construction of the gap instance for \scut\ as follows.

We modify the gap instance we got for \balsep to get an instance for the linear round Lasserre relaxation of \scut. The reduction converts the gap instance for \balsep to the gap instance for \scut in an almost black box style. In the \balsep problem, we have the hard constraint that the cut is $\tau$-balanced. In the reduction from \balsep to \scut, we need to use the sparsity objective to enforce this constraint. We do it as follows. Recall that given a \threexor instance $\Phi$, the corresponding gap instance for \balsep consists of vertex set $L_\Phi \cup R_\Phi \cup Z_r$ and edge set $\mathcal{E}_\Phi$. To get a gap instance for \scut, we add two more $O(M)$-regular expanders (with edge expansion $10^4 \cdot M$) $D_l$ and $D_r$ of size $1000 M m$ (where $M$ is the same parameter defined in the previous sections). Now,  let the edge set $\mathcal{E}'_\Phi$ contain the edges in $\mathcal{E}_\Phi$, in the expanders $D_l$ and $D_r$, and the following edges : for each vertex $v \in L_\Phi \cup R_\Phi \cup Z_r$, introduce $2$ new edges incident to it, one to a vertex in $D_l$ (say, $v_l$) and the other one to a vertex in $D_r$ (say, $v_r$). We arrange these edges (between $L_\Phi \cup R_\Phi \cup Z_r$ and $D_l$, $D_r$) in a way so that each vertex in $D_l$ (or $D_r$) has at most one neighbor in $L_\Phi \cup R_\Phi \cup Z_r$ -- this can be done because $|L_\Phi| + |R_\Phi| + |Z_r| = (2M + 5)m < 1000Mm = |D_l| = |D_r|$.

Using the instance described above, we will prove our main integrality gap theorem (\pref{thm:informal-scut}) for \scut. We state the full theorem as follows.
%
%\deferapx{apx:2}
%\newcommand{\scutthm}
\begin{theorem}\label{thm:scutig}
{For large enough constants $\beta, M$ (where $\beta$ is the same parameter as in previous sections), and infinitely many positive integer $N$'s, there is an $N$-vertex instance for \scut problem, such that the optimal solution is at least $(1 + 1/(100M))$ times worse than the optimal solution of the $\Omega(N)$-round Lasserre SDP.}
\end{theorem}
%\newcommand{\scutproof}
%Due to space constraints, the proof is deferred to \pref{app:scut}.

\pref{thm:scutig} is directly implied by the following completeness lemma (\pref{lem:scut-completeness}) and soundness lemma (\pref{lem:scut-soundness}).

\begin{lemma}[Completeness]\label{lem:scut-completeness}
The value of relaxation $\Psi_2$ (in \pref{fig:scut}) is at most  \[ (2M + 10)m / ((1001M + 1)m)^2 \] for $\tau = (1001M + 1)/(2002M + 5)$.
\end{lemma}
\begin{proof}
%{
%\paragraph{Completeness.}
Given the SDP solution  $\{\bbU_{S'}\}_{S' \subseteq L_\Phi \cup R_\Phi \cup Z_r, |S'| \leq r/3}$ in the completeness case of \balsep, we extend it to the SDP solution  $\{\bbU_{S}\}_{S \subseteq L_\Phi \cup R_\Phi \cup Z_r \cup D_l \cup D_r, |S| \leq r/3}$  for \scut  by ``putting $D_l$ and $D_r$ one per side". That is, for each $S \subseteq L_\Phi \cup R_\Phi \cup Z_r \cup D_l \cup D_r$ with $|S| \leq r/3$, let $S' = S \cap ( L_\Phi \cup R_\Phi \cup Z_r)$. Now we let $\bbU_S = \bm{0}$ if $S \cap D_r \neq \emptyset$, and let $\bbU_S = \bbU_{S'}$ otherwise.

We first check that $\{\bbU_{S}\}_{S \subseteq L_\Phi \cup R_\Phi \cup Z_r \cup D_l \cup D_r, |S| \leq r/3}$ is a feasible SDP solution. We only check that the balance constraint (the last constraint in relaxation $\Phi_2$) is met.

We are going to prove prove that
\[
\sum_{u \in L_\Phi \cup R_\Phi \cup Z_r \cup D_l \cup D_r} \bbU_{\{u\}} = (1001M + 1)m \bbU_{\emptyset} .
\]

From the proof of \pref{lem:bs-completeness}, we know that
\[
\sum_{u \in L_\Phi \cup R_\Phi \cup Z_r} \bbU_{\{u\}} = (M + 1) m \bbU_{\emptyset} ,
\]
together with the fact that
\begin{align*}
\forall u \in D_l, \bbU_{\{u\}} = \bbU_{\emptyset}, \qquad \forall u \in D_r, \bbU_{\{u\}} = \bm{0},
\end{align*}
we get the desired equality.

Now we calculate the value of the solution. First, we calculate the following value.

\begin{align*}
&\sum_{(u, v)\in \mathcal{E}'_\Phi}  \norm{\bbU_{\{u\}} - \bbU_{\{v\}}}^2 =\sum_{(u, v)\in \mathcal{E}_\Phi}  \norm{\bbU_{\{u\}} - \bbU_{\{v\}}}^2 + \sum_{(u, v)\in \mathcal{E}'_\Phi \setminus \mathcal{E}_\Phi}  \norm{\bbU_{\{u\}} - \bbU_{\{v\}}}^2\\
=& 5m + \sum_{u, v \in D_l}\norm{\bbU_{\{u\}} - \bbU_{\{v\}}}^2 + \sum_{u, v \in D_r}\norm{\bbU_{\{u\}} - \bbU_{\{v\}}}^2 \\
& \qquad \qquad + \sum_{u \in L_\Phi \cup R_\Phi \cup Z_r} \left (\norm{\bbU_{\{u\}} - \bbU_{\{v_l\}}}^2 + \norm{\bbU_{\{u\}} - \bbU_{\{v_r\}}}^2 \right),
\end{align*}
Note that $ \sum_{u, v \in D_l}\norm{\bbU_{\{u\}} - \bbU_{\{v\}}}^2 + \sum_{u, v \in D_r}\norm{\bbU_{\{u\}} - \bbU_{\{v\}}}^2 = 0$, and
\begin{align*}
 &\sum_{u \in L_\Phi \cup R_\Phi \cup Z_r}\left (\norm{\bbU_{\{u\}} - \bbU_{\{v_l\}}}^2 + \norm{\bbU_{\{u\}} - \bbU_{\{v_r\}}}^2 \right) \\
=&\sum_{u \in L_\Phi \cup R_\Phi \cup Z_r}\left( 2\norm{\bbU_{\{u\}}}^2 + \norm{\bbU_{\{v_l\}}}^2 + \norm{\bbU_{\{v_r\}}}^2 - 2 \langle \bbU_{\{u\}}, \bbU_{\{v_l\}} \rangle - 2 \langle \bbU_{\{u\}}, \bbU_{\{v_r\}}\rangle \right)\\
=&\sum_{u \in L_\Phi \cup R_\Phi \cup Z_r}\left( 2\norm{\bbU_{\{u\}}}^2 + 1 + 0 - 2 \norm{\bbU_{\{u, v_l\}}}^2 - 2\norm{ \bbU_{\{u, v_r\}} }^2 \right) \qquad \text{(by property of Lasserre vectors)}\\
=&\sum_{u \in L_\Phi \cup R_\Phi \cup Z_r} 1 = |L_\Phi| + |R_\Phi| + |Z_r| = (2M + 5) m .
\end{align*}
Thus, we have
\begin{align*}
\sum_{(u, v)\in \mathcal{E}'_\Phi}  \norm{\bbU_{\{u\}} - \bbU_{\{v\}}}^2 = (2M + 10) m .
\end{align*}
Since $\tau < 1/2$, the value of the solution is at most
\begin{align*}
\frac{1}{|L_\Phi \cup R_\Phi \cup Z_r \cup D_l \cup D_r|^2 \tau^2} \sum_{(u, v)\in \mathcal{E}'_\Phi}  \norm{\bbU_{\{u\}} - \bbU_{\{v\}}}^2 = (2M + 10) m /((1001M+1)m)^2.
\end{align*}

\iffalse
Now, we estimate the denominator.
\begin{align*}
&\frac{1}{2} \sum_{u, v \in  L_\Phi \cup R_\Phi \cup Z_r \cup D_l \cup D_r}  \norm{\bbU_{\{u\}} - \bbU_{\{v\}}}^2\\
& \geq \frac{1}{2} \sum_{u, v \in  R_\Phi} \norm{\bbU_{\{u\}} - \bbU_{\{v\}}}^2 + \sum_{u \in D_l, v \in D_r} \norm{\bbU_{\{u\}} - \bbU_{\{v\}}}^2 + \sum_{u \in R_\Phi, v \in D_l \cup D_r}\norm{\bbU_{\{u\}} - \bbU_{\{v\}}}^2 .
\end{align*}

We know that
\begin{align*}
& \frac{1}{2} \sum_{u, v \in  R_\Phi} \norm{\bbU_{\{u\}} - \bbU_{\{v\}}}^2\\
=& \frac{(M\beta)^2}{2} \sum_{j, j'} \sum_{ \alpha, \alpha'} \norm{\bbU_{\{(x_j, \alpha, 1)\}} - \bbU_{\{(x_{j'}, \alpha', 1)\}}}^2 \\
=& \frac{(M\beta)^2}{2} \left(2n \sum_{j} \sum_{\alpha} \norm{\bbU_{\{(x_j, \alpha, 1)\}}}^2 - 2 \sum_{j, j'} \left\langle \sum_{ \alpha} \bbU_{\{(x_j, \alpha, 1)\}} , \sum_{\alpha'} \bbU_{\{(x_{j'}, \alpha', 1)\}}\right\rangle  \right) \\
\geq& \frac{(M\beta)^2}{2} (4n^2 - 2n^2) = (M\beta n)^2 = (Mm)^2 ,
\end{align*}
and
\begin{align*}
\sum_{u \in D_l, v \in D_r} \norm{\bbU_{\{u\}} - \bbU_{\{v\}}}^2 = \sum_{u \in D_l, v \in D_r} \norm{\bbU_{\{u\}} }^2  = (1000 Mm)^2,
\end{align*}
and
\begin{align*}
\sum_{u \in R_\Phi, v \in D_l \cup D_r}\norm{\bbU_{\{u\}} - \bbU_{\{v\}}}^2 = \sum_{u \in R_\Phi, v \in D_l} (1 - \norm{\bbU_{\{u\}}}^2) + \sum_{u \in R_\Phi, v \in D_r}  \norm{\bbU_{\{u\}}}^2 = 2 \cdot 1000 (Mm)^2 .
\end{align*}

Therefore, the denominator is lower-bounded by $(1001Mm)^2$, and we conclude that the SDP value is at most $(2M + 9 + \sqrt{M/m})m / (1001Mm)^2$ .

\fi
\end{proof}

\begin{lemma}[Soundness]\label{lem:scut-soundness}
For large enough $M$, the sparsity of the sparsest cut is at least $\gamma = (1 + 1/(100M)) \cdot(2M + 10) m /(1001Mm)^2$.
\end{lemma}
\begin{proof}
%\paragraph{Soundness.}
Let $D_l'$ be the smaller part among $D_l \cap S$ and $D_l \cap \bar{S}$, and $D_l''$ be the larger part. Also, let $D_r'$ be the smaller part among $D_r \cap S$ and $D_r \cap \bar{S}$ and $D_r''$ be the larger part. Let $(T, \bar{T})$ be the cut restricted to $L_\Phi \cup R_\Phi \cup Z_r$ (the \balsep  instance), i.e. let $T = S \cap (L_\Phi \cup R_\Phi \cup Z_r)$ and $\bar{T} = \bar{S} \cap (L_\Phi \cup R_\Phi \cup Z_r)$.

First, we show that to get a cut of sparsity better than $\gamma$, $|D_l'| \leq \frac{1}{10^4 M} \cdot |D_l|$, and the same is true for $D_r$ (by the same argument). This is because if $|D_l'| > \frac{1}{10^4 M} \cdot |D_l|$, by the expansion property, there are at least $10^4 M \cdot |D_l'| > 1000Mm$ edges in the cut. Since the graph has $|L_\Phi| + |R_\Phi| + |Z_r| + |D_l| + |D_r| = (2002M + 5)m$ vertices, therefore the sparsity of the cut is at least
\begin{align*}
\frac{1000Mm}{\frac{1}{4} \cdot ((2002M + 5)m)^2} > \frac{500Mm}{(1001Mm)^2} > \gamma,
\end{align*}
for $M > 1/25$.

Second, we show that $D_l'$ and $D_r'$ should be on opposite sides of any cut of sparsity better than $\gamma$. Suppose not, let $S$ be the side of the cut which $D_l'$ and $D_r'$ are on. Recall that $T = S \cap (L_{\Phi} \cup R_{\Phi} \cup Z_r)$. We have
\begin{align*}
\edges(S, \bar{S}) \geq \edges(T, D_l'' \cup D_r'') + \edges(D_l', D_l'') + \edges(D_r', D_r'') .
\end{align*}
Note that $\edges(T, D_l'' \cup D_r'') \geq 2|T| - |D_l'| - |D_r'|$ as each vertex in $D_l, D_r$ is connected to at most one vertex in $T$. Also, by the expansion property,  $\edges(D_l', D_l'') + \edges(D_r', D_r'') \geq 1000M (|D_l'| + |D_r'|)$. Now, we have
\begin{align*}
\edges(S, \bar{S}) &\geq 2|T| - (|D_l'| + |D_r'|) + 1000M (|D_l'| + |D_r'|) \\
&= 2(|T| + |D_l'| + |D_r'|)  + (1000M - 3)(|D_l'| + |D_r'|) \geq  2(|T| + |D_l'| + |D_r'|)  = 2|S| .
\end{align*}

Therefore, the sparsity of the cut
\begin{align*}
\frac{\edges(S, \bar{S})}{|S||\bar{S}|} &\geq \frac{2|S|}{|S| |\bar{S}|} = \frac{2}{|\bar{S}|} \geq \frac{2}{(2002M + 5)m} > \gamma .
\end{align*}

Third, we show that if the cut $(S, \bar{S})$ has sparsity better than $\gamma$, then the cut $(T, \bar{T})$ defined above is a $0.49$ vs $0.51$ balanced cut, i.e. $|T| / (|L_\Phi| + |R_\Phi| + |Z_r|) \in [0.49, 0.51]$. Supposing $(T, \bar{T})$ is not $0.49$ vs $0.51$ balanced, i.e. $||T| - |\bar{T}|| > 0.02 \cdot (2M + 5)m$, we have
\begin{align*}
||S| - |\bar{S}|| \geq ||T| - |\bar{T}|| - |D_l'| - |D_r'| & \geq 0.02 \cdot (2M + 5)m - \frac{2}{1000M} \cdot 1000 M m \\
& \geq (0.04 M - 2)m \geq 0.01 M m,
\end{align*}
for large enough $M$. Therefore, $(S, \bar{S})$ is not $0.5 - 10^{-6}$ vs $0.5 + 10^{-6}$ balanced. Thus,
\begin{align*}
|S| |\bar{S}| < ((2002 M + 5)m)^2 \cdot (0.5 - 10^{-6})(0.5 + 10^{-6}) < (1001 Mm)^2\cdot (1 - 10^{-12}) .
\end{align*}
Since $D_l''$ and $D_r''$ are on opposite sides of $(S, \bar{S})$, we know that $\edges(S, \bar{S}) \geq (2M + 5)m - |D_l'| - |D_r'| \geq (2M + 5)m \cdot (1 - 1/M)$, and therefore the sparsity of the cut
\begin{align*}
\frac{\edges(S, \bar{S})}{|S| |\bar{S}|} > \frac{(2M + 5)m}{(1001Mm)^2} \cdot (1 - 1 / M) (1 + 10^{-12}).
\end{align*}
This value is greater than $\gamma$ when $M > 10^{20}$.

Finally, since $(T, \bar{T})$ is a $0.49$ vs $0.51$ balanced cut, by \pref{lem:bs-soundness}, we know that with probability $1 - o(1)$, $\edges(T, \bar{T}) > (5.4 - O(1 / \sqrt{\beta}) - O(1/M)))m$. Therefore
\begin{align*}
&\frac{\edges(S, \bar{S})}{|S| |\bar{S}|} \\
\geq & \frac{\edges(T, \bar{T}) + (2M + 5)m - |D_l'| - |D_r'|}{\frac{1}{4}\cdot ((2002M + 5)m)^2} \\
\geq & \frac{(5.4 - O(1 / \sqrt{\beta}) - O(1/M)))m + (2M + 5)m - \frac{1000Mm}{10^4 M} - \frac{1000Mm}{10^4 M}}{\frac{1}{4}\cdot ((2002M + 5)m)^2} \\
=& \frac{(2M + 10.2  - O(1/\sqrt{\beta}) - O(1/M))m}{\frac{1}{4}\cdot ((2002M + 5)m)^2}\\
\geq& \frac{(2M + 10.2  - O(1/\sqrt{\beta}) - O(1/M))m}{(1001Mm)^2} \cdot (1 - 1/(200M)) \\
\geq& \frac{(2M + 10.1 )m}{(1001Mm)^2} \cdot (1 - 1/(200M)) & \text{(for large enough $\beta$ and $M$)}\\
\geq& \frac{(2M + 10)m}{(1001Mm)^2} \cdot (1 + 1/(30M))(1 - 1/(200M))  & \text{(for large enough $M$)}\\
\geq& \frac{(2M + 10)m}{(1001Mm)^2} \cdot (1 + 1/(100M)) = \gamma .
\end{align*}
\end{proof}

\bibliographystyle{alpha}

\newcommand{\etalchar}[1]{$^{#1}$}

\end{document}